\documentclass{article}

\usepackage{arxiv}
\usepackage{graphicx, amsmath, amsfonts, setspace, hyperref}
\usepackage{algorithm}
\usepackage{algpseudocode}
\usepackage{booktabs, multirow}
\usepackage[flushleft]{threeparttable}
\usepackage{changepage} 
\usepackage{booktabs,caption}

\title{Exploring Supervised Machine Learning for Multi-Phase Identification and Quantification from Powder X-Ray Diffraction Spectra}

\author{
	\href{https://orcid.org/0000-0001-7821-6053}{\includegraphics[scale=0.06]{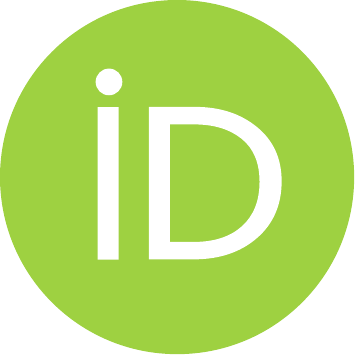}\hspace{1mm}Jaimie Greasley} \\
	Department of Physics\\
	The University of the West Indies\\
	St. Augustine, Trinidad\\
	\texttt{jaimie.greasley@gmail.com} \\
	\And
	\href{https://orcid.org/0000-0003-1729-559X}{\includegraphics[scale=0.06]{orcid.pdf}\hspace{1mm}Patrick Hosein}\\
	Department of Computer Science\\
	The University of the West Indies\\
	St. Augustine, Trinidad\\
	\texttt{patrick.hosein@sta.uwi.edu} \\
}

\renewcommand{\shorttitle}{Supervised Machine Learning for PXRD Multi-Phase Identification and Quantification}

\hypersetup{
pdftitle={Exploring Supervised Machine Learning for Multi-Phase Identification and Quantification from Powder X-Ray Diffraction Spectra},
pdfauthor={Jaimie Greasley and Patrick Hosein},
colorlinks=true
}

\begin{document}

\maketitle

\begin{abstract}
Powder X-ray diffraction analysis is a critical component of materials characterization methodologies. Discerning characteristic Bragg intensity peaks and assigning them to known crystalline phases is the first qualitative step of evaluating diffraction spectra. Subsequent to phase identification, Rietveld refinement may be employed to extract the abundance of quantitative, material-specific parameters hidden within powder data. These characterization procedures are yet time-consuming and inhibit efficiency in materials science workflows. The ever-increasing popularity and propulsion of data science techniques has provided an obvious solution on the course towards materials analysis automation. Deep learning has become a prime focus for predicting crystallographic parameters and features from X-ray spectra. However, the infeasibility of curating large, well-labelled experimental datasets means that one must resort to a large number of theoretic simulations for powder data augmentation to effectively train deep models. Herein, we are interested in conventional supervised learning algorithms in lieu of deep learning for multi-label crystalline phase identification and quantitative phase analysis for a biomedical application. First, models were trained using very limited experimental data. Further, we incorporated simulated XRD data to assess model generalizability as well as the efficacy of simulation-based training for predictive analysis in a real-world X-ray diffraction application.

\end{abstract}

\section*{Introduction}

The acquisition and assessment of X-ray diffraction (XRD) spectra is the central tenet of many materials characterization investigations. For a material of interest, x-ray scattering data reveals structural and microstructural parameters in addition to essential information on the intrinsic symmetrical arrangement of atoms in lattices \cite{dinnebier2008powder}. The collection of scattered intensities corresponding to a set of interplanar \textit{d}-spacing magnitudes serves as a fingerprint reference for the material structure \cite{pecharsky2009fundamentals}. Justifiably, the diffraction spectra of powdered materials are compiled into large databases consisting up to hundreds of thousands of records for experimental reference \cite{gates2019powder}.

The task of matching diffraction peaks to known crystalline phases is not without effort. Regardless the use of search-match software, in the absence of contextual information on the nature of the sample or a great deal of expertise, phase identification  is far from automatic \cite{lutterotti2019full}. Differing instrument settings and sample-specific features produce altered versions of the expected diffraction profile for a given crystalline phase. This obscures the process of cross-referencing against the database records. Phase matching is further obstructed when dealing with powder patterns of poorly crystallized or multi-phase materials due to exacerbated peak overlap. It is time-inefficient, bordering on impractical, to sieve through the numerous search-match hits for visual assessment of each record, and then to fill in the blanks via trial-and-error to account for unassigned peaks. Moreover, wherever a complete characterization of the sample is needed, it is vital to perform a whole powder pattern fitting (WPPF) routine like the Rietveld method \cite{rietveld1967line}. The latter permits the extraction of refined material parameters including phase fractions, lattice dimensions, atom positions, occupancy factors, crystallite size, texture and strain. Despite the magnitude of data retrieved, Rietveld refinement from powder x-ray data is onerous, sometimes requiring an hour or more to complete\cite{mccusker1999rietveld,oviedo2019fast}. 

Attention has already been drawn to such impediments in materials characterization, and with particular concern in the context of high-throughput experimentation (HTE). For instance in thin-films, the high research capacity gained by state-of-the-art methods to rapidly synthesize and screen large combinatorial libraries is bottle-necked at the stage of data analysis \cite{mao2015combinatorial,sun2019accelerated}. However, the genesis of exploring machine learning (ML) for materials analysis automation began as a result of the inclination to overcome such problems. In the same manner, the launching of the Materials Genome Initiative (MGI) \cite{de2019new} has propelled an ease-of-access to both synthetic and experimental datasets within the last decade. These foundations have culminated with an explosion of data-driven research in the materials science domain with the evolution of  materials informatics, also deemed ``the fourth paradigm" \cite{agrawal2019deep}. 

Within the framework of supervised machine learning algorithms, exceptional emphasis has been placed so far on deep learning (DL) for predicting various material properties from crystallographic data \cite{choudhary2022recent}. In the line of X-ray diffraction profile analysis, work has been done on crystal system prediction \cite{park2017classification, vecsei2019neural,aguiar2020crystallographic,zaloga2020crystal}, space group determination \cite{park2017classification,oviedo2019fast,vecsei2019neural} as well as indicating the presence of various crystalline phases \cite{lee2020deep,wang2020rapid,lee2021data,maffettone2021crystallography,szymanski2021probabilistic}. For phase classification, Lee et al. \cite{lee2020deep} used a convolutional neural network (CNN) to discern 38 inorganic phases within a quaternary system. The model was trained with about 1.8 million simulations. Similarly, Szymanski et al. \cite{szymanski2021probabilistic} executed  an ensemble CNN trained with around 38 thousand synthetic profiles for phase identification in monophasic and multi-phasic compositions. The estimation of quantitative structural and compositional parameters is the alternate objective of deep learning research for diffraction data analysis. CNNs have been employed also in the interest of lattice parameter \cite{chitturi2021automated,dong2021deep}, crystallite size \cite{dong2021deep} and phase fraction prediction \cite{lee2020deep,lee2021data}. 
 
A deep learning approach can afford high accuracy without a requirement of data preprocessing or feature engineering. Yet, DL is also criticized for lack of interpretability due to an inherently complex construction which renders it difficult to extract contextual rule-based knowledge \cite{agrawal2019deep}. The greater drawback for deep learners is however the considerable number of training parameters which call for labelled `big' data and a substantial amount of time for any effective training of the model. In many scenarios, well-characterized `big' XRD data is only conceivable by supplementing experimental spectra with simulated data.

Maffettone et al. \cite{maffettone2021crystallography} described an ensemble system composed of 50 CNNs for phase identification. Whilst the model does not seek explicitly labelled input data from the user, it requires hours to construct and train the ensemble with synthetically generated spectra from a pre-defined phase library. One study by Lee et al. \cite{lee2021data} sought to boost the previous report's CNN accuracy  \cite{lee2020deep} for phase fraction prediction. They investigated for 21 compounds of a quartenary system relevant to solid-state electrolytes. Oddly enough, after augmenting the dataset to nearly 14 million simulations, the authors still reported poor performance for deep network architectures and concluded that a single hidden-layer neural network gave higher accuracy. Better results were even seen with conventional learning models like random forest (RF), $k$-nearest neighbors (k-NN) and the support vector machine (SVM).    

On reviewing the existing literature, it has become evident that simpler learning models are often overlooked without reasonable domain-specific or empirical justification. Few studies \cite{bunn2015generalized,park2022application} have endeavoured first to investigate elementary supervised learning algorithms like nearest neighbors, support vector machines, decision trees, naive Bayes and ensemble techniques for multi-label phase identification of XRD data. Models like these are formulated by hypothesizing some expected property or structure to the data prior to training whereas neural networks attempt to learn the structure entirely, thereby requiring many more training examples. In the present paper, we have taken an interest in evaluating the application of elementary supervised learning algorithms rather than deep learners to the task of predicting mineral phase composition for a medical application. What is also relevant to our aim is estimating the relative phase weights in multi-phasic compositions. We have applied several regression models to this task and evaluated their performance. Below is an outline of the biomedical application context. 

\subsection*{Kidney Stone Analysis}
Stone disease is an intensely painful condition with prevalence falling usually in the 5-10\% range for a population \cite{qian2022epidemiological}. Kidney stones or more precisely urinary tract calculi, arise from the nucleation, growth and clustering of mineralogical crystals propelled by persistent supersaturation of constituent ions in urine. Kidney stones often exhibit a high degree of crystallinity \cite{mirkovic2020phase}, and are composed of mainly inorganic phases such as calcium-based oxalates (CaOx) and phosphates (CaPh) as well as magnesium-based phosphates. The less common stone compositions are uric acid, urates and protein phases.

The formation of calculi in the urinary tract is an aberrant event catalyzed by some anatomical, genetic, metabolic, dietary or environmental anomaly for the individual \cite{daudon2016comprehensive}. Resolving the exact composition of a stone often gives strong indicators to the pathology driving its formation. This guides medical practitioners in diagnosing the underlying condition and deriving a target-specific treatment plan. Given a first-time stone event, there is an elevated risk for another stone. The presence of specific mineral phases are also pointers to the likeliness and degree of recurrent disease. 
In Table \ref{tab:ks}, some common kidney stone minerals, their frequencies, associated pathologies and risk for recurrence are outlined. 

For the reasons outlined above, stone analysis is highly recommended for all first-time patients by the major international urological associations \cite{pearle2014medical,Turk_EAU_2020}. Moreover, stone analysis via powder x-ray diffraction is by far the most accurate and sophisticated method for the precise identification of mineral phase composition in kidney stones. Recently, we reported XRD Rietveld characterization of mineral phases in a batch of 46 urinary tract stones \cite{greasley2022quantitative}. We identified 7 distinct phases throughout the study, with an average of 2.2 phases being detected in each stone. The maximum number identified was 4 phases per sample which was the case for 13\% of the batch. Given that quantitative Rietveld analysis becomes increasingly laborious and time-consuming with each additional phase, it is warranted that we have sought an automated approach for a routine stone analysis program. Here, we have furthered our investigation to evaluate the performance of supervised learning models for identifying and quantifying mineral phases from powder x-ray diffraction spectra in context of the described application. 

\begin{table}[h]
\centering
\caption{Frequency, Recurrence Risk and Pathological Associations for Minerals found in Kidney Stones \cite{schubert2006stone,daudon2016comprehensive,daudon2018recurrence}}
\label{tab:ks}
\vspace{2mm}
\resizebox{\textwidth}{!}{%
\begin{tabular}{llll}
\hline
\textbf{Mineral Phase} &
  \textbf{Frequency} &
  \textbf{Risk Level} &
  \textbf{Common Associations} \\ \hline
\textit{Calcium oxalate monohydrate (Whewellite)} &
  78\% &
  \begin{tabular}[c]{@{}l@{}}typically low, \\ except for primary hyperoxaluria\end{tabular} &
  \begin{tabular}[c]{@{}l@{}}primary hyperoxaluria, secondary hyperoxaluria, Randall's plaque,\\  inflammatory bowel disease, chronic pancreatitis\end{tabular} \\
\textit{Calcium oxalate dihydrate (Weddellite)} &
  48\% &
  low &
  hypercalciuria, hypocitraturia, primary hyperthyroidism \\
\textit{Carbonated hydroxyapatite} &
  33\% &
  \begin{tabular}[c]{@{}l@{}}low to medium,\\ high for dRTA\end{tabular} &
  \begin{tabular}[c]{@{}l@{}}distal renal tubular acidosis (dRTA), hypercalciuria, \\ urinary tract infection (UTI)\end{tabular} \\
\textit{Magnesium ammonium phosphate hexahydrate (Struvite)} &
  6\% &
  medium to high &
  urinary tract infection (UTI) by urea splitting organisms \\
\textit{Uric acid (Uricite)} &
  10\% &
  medium to high &
  \begin{tabular}[c]{@{}l@{}}low urine pH, insulin resistance, type II diabetes, metabolic syndrome,\\ morbid obesity\end{tabular} \\
\textit{Ammonium acid urate} &
  1\% &
  medium to high &
  hyperuricusoria, urinary tract infection, chronic diarrhea, laxative abuse \\
\textit{Calcium hydrogen phosphate dihydrate (Brushite)} &
  1-2\% &
  high &
  hypercalciuria, primary hyperthyroidism, recurrent stones \\
\textit{Cystine} &
  \textless{}1\% &
  high &
  cystinuria \\ \hline
\end{tabular}%
}
\end{table}

\section*{Experimental Methods}
\subsection*{Data Preparation}
Urinary tract calculi were previously collected from consenting adult patients at public health institutions. Powder x-ray diffraction spectra were measured with a Bruker D2 Phaser bench-top diffractometer for an angular range $2^{\circ}\leq 2\theta \leq 55^{\circ}$ at step-size of 0.02$^{\circ}$. Preliminary phase identification was performed by means of search/match in DIFFRAC.EVA (v4.2) utilizing the International Center for Diffraction Data (ICDD) PDF-2 database. Whole profile fitting was achieved with MAUD \cite{lutterotti1999maud} Rietveld refinement software for verification of phase presence as well as for obtaining optimized sample parameters including weight fractions. 

\begin{figure}[h]
\centering
\includegraphics[width=0.9\linewidth]{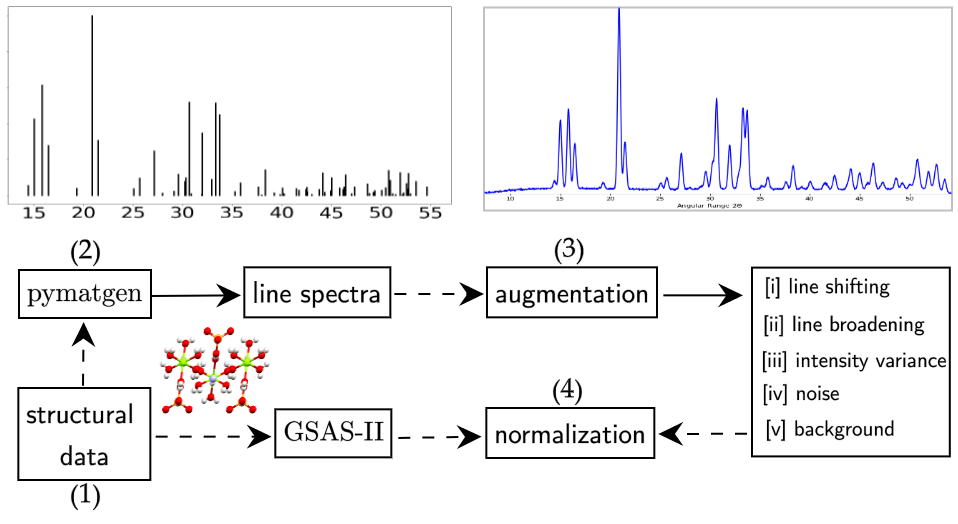}
\caption{A schematic of the (a) the powder x-ray diffraction spectra simulation and experimental process and (b) the formulated supervised learning model for phase fraction estimation}
\label{fig:scheme}
\end{figure}

Synthetic XRD patterns were prepared with a variety of methods. The theoretical line spectra of those phases which we previously found to have larger crystallites were initiated with \textit{pymatgen} middleware \cite{ong2013python} using the relevant structural CIF files obtained from the Crystallography Open Database \cite{gravzulis2009crystallography}. The physics-based data augmentation protocol by Szymanski et al.\cite{szymanski2021probabilistic} was then applied for introducing zero offset, non-uniform lattice parameter and strain-related variance of peak positions, line spectra broadening due to finite crystallite sizes and modified peak intensity signals arising from preferred orientation. The addition of a 4th to 7th order background polynomial with random noise was another experimental effect we incorporated. It was also necessary to replicate the broadened overlapping peaks that were observed experimentally for phases with smaller than usual and even nanometric crystallite dimensions. For these phases, GSAS-II software \cite{toby2013gsas} was utilized for realistic XRD simulation. Multi-phase patterns were fabricated by linearly combining weighted monophasic spectra. The different data collection and preparation methods meant that we were unable to select appropriate scale factors for the simulated data. For consistency, both the experimental and simulated datasets were normalized relative to the maximum signal value of each spectrum. Figure \ref{fig:scheme} summarizes the data preparation method. 

\subsection*{Supervised Models}
We investigated several supervised learning models and a couple ensemble methods for multi-label phase classification: $k$-Nearest Neighbours (\textbf{k-NN}), Support Vector Machine (\textbf{SVM}), Decision Tree (\textbf{DTr}), Gaussian Naive Bayes (\textbf{GNB}), Multinomial Naive Bayes (\textbf{MNB}), Complement Naive Bayes (\textbf{CNB}), Random Forest (\textbf{RF}) and Extremely Randomized Trees (\textbf{ExTr}). The optimized hyper-parameters for each model were selected according to the Scikit-learn \cite{pedregosa2011scikit} GridSearchCV method tested on randomized sets of the data. For comparison, we also trained a shallow artificial neural network (\textbf{ANN}). The ANN comprised 1000 nodes in its single hidden layer with logistic function activation, and used the ADAM optimization algorithm for training. Other network architectures were initially tried on simulated data by varying the number of nodes in the hidden layer by increments of 100. We also tested one double hidden layer network structure. With 5-fold cross-validation, the performances of the single HL networks were relatively similar, with the 1000-node ANN giving only a slightly better average overall. We persisted with this construction for the rest of the study. For the second objective of predicting the weight fractions of crystalline phase labels, several of the above classification algorithms like tree-based methods and nearest neighbors, already are inherently structured for multi-output regression. We were able to test equivalent multi-target regression versions for the k-NN, SVM, DTr, RF and ExTr algorithms as well as ANN.
 
\subsection*{Evaluation Metrics}
Regarding the multi-label classification problem, we utilized the standard metrics like accuracy, precision and recall. Accuracy is defined in the conventional manner as the sum of true positive (TP) and true negative (TN) predictions divided by the total number of predictions. Precision and recall take into account false positive and false negative predictions respectively. A high false positive (FP) rate is homologous with low precision whereas a high false negative (FN) rate correlates to low recall. There is typically a trade-off made in assessing model performance with these two metrics as modification of various hyper-parameters to improve one often results in a declined performance for the other. A given application context may emphasize the relative importance of high precision versus that of high recall required of the predictive model. In stone analysis, the inability to recognize certain high-risk phases even in small quantities, can lead to an inaccurate diagnosis of the problem and inadequate treatment for recurrence prevention. Yet, for the more common phases that carry lower risk for recurrence and complications, a false negative prediction may not necessarily modify the overall conclusion of the analysis. On the other hand, a model that is not very precise is practically ineffective. 

The $F_1$ score metric places equal importance on precision and recall [eq. (\ref{eq:f1})]. There are several methods for calculation, namely `macro', `micro', `weighted' and `samples'. Consider $T$ testing samples and the ascribed $K$ labels. The micro $F_1$ method relates to the total precision and recall of the $T\times K$ predictions whilst macro $F_1$ considers precision and recall for each of the $K$ labels singly then takes the average. In the case of an imbalanced dataset, the weighted $F_1$ score is preferred to the macro as it gives more credit to the precision-recall values of the better represented labels in the dataset. The samples $F_1$ score is applicable to multi-label problems where each sample can be predicted positive for more than one of the $K$ classes. The precision-recall value for each sample is evaluated and the average is taken across the $T$ samples. 

\begin{equation}
    F_1= 2\times\frac{\text{precision}\cdot\text{recall}}{\text{precision}+\text{recall}}
    \label{eq:f1}
\end{equation}

For quantitative phase estimations, three regression metrics were applied for performance evaluation. Considering a single test sample, we denote the vectors of the true and predicted phase fractions as  $\alpha$ and $\hat{\alpha}$ respectively. The mean absolute error (MAE) [eq. (\ref{eq:mae})] enables a direct quantification of the average difference between the real and projected values for each prediction made by the model.
\begin{equation}
    \textbf{MAE}=\frac{1}{TK} \sum_{s,j} \left|\alpha_{s,j}-\hat{\alpha}_{s,j}\right|
    \label{eq:mae}
\end{equation}

The cosine similarity metric (CS) measures the extent to which two vectors point in the same direction, irrespective of their norm magnitudes. The cosine similarity is expressed as in eq. (\ref{eq:sc}) below.

\begin{equation}
    \textbf{CS} \equiv \cos{\theta}= \frac{\alpha\cdot\hat{\alpha}}{||\alpha||\times||\hat{\alpha}||}
    \label{eq:sc}
\end{equation}

For our application, CS may be useful for outlining similarity in the relative ratios between predicted phases rather than their exactly predicted quantities. Two vectors $\vec{a}=[1,2]$ and $\vec{b}=[4,8]$ give a perfect cosine similarity of 1 as the relative ratio of the second vector component to the first is 2:1, i.e. their slopes are equal.

We now define the performance metric $\rho$ [eq. (\ref{eq:rho})] which accounts for both the magnitudes and relative directions of the predicted and true label vectors. The Euclidean distance between the two vectors is represented as $||\alpha - \hat{\alpha}||_2$. As the elements in each vector $\alpha$ and $\hat{\alpha}$ must sum to 1, the maximum value of the Euclidean distance is $\sqrt{2}$, the case of orthogonal vectors.

\begin{equation}
\rho = 1 - \frac{||\alpha - \hat{\alpha}||_2}{\sqrt{2}}
\label{eq:rho}
\end{equation}

If therefore $\alpha = \hat{\alpha}$, we get $\rho=1$ which implies a perfect prediction. If there is only one phase and the model predicts any other phase except the true phase, then $\rho=0$ indicating a wrong prediction. Considering now a bi-phasic example where two phases are correctly predicted but the true fractions are 0.5 and 0.5 and the predicted fractions are 0.6 and 0.4. We obtain $\rho=0.9$ indicating a close but not perfect prediction. Whilst the $\rho$ metric is ideal for the problem, regression algorithms are not explicitly constrained to maintain positive, normalized outputs, though this may be the expectation given the input training data. 

\section*{Results \& Analysis}
\subsection*{Phase Identification - Multi-Label Classification}
First was to evaluate whether traditional learning algorithms are useful for indicating the presence of crystalline phases from experimental powder data. Supervised learning models selected for this part of the study were those already adapted for multi-label function using continuous-variable features in the Scikit-learn library. These were trained until convergence prior to testing. We carried out four instances of assessment: [1] simulation training with simulation testing on the primary and secondary simulated datasets, [2] experimental training with experimental testing, [3] simulation training using the primary simulated dataset with experimental testing and lastly [4] training using the primary simulated dataset and randomly selected experimental examples followed by testing with the remaining experimental data.  Our primary simulated dataset consisted of 105 monophasic spectra augmented from the 7 crystalline phases that were identified for the stone batch previously \cite{greasley2022quantitative}. The secondary dataset (N=184) comprised the original in addition to several weighted linear combinations of the primary spectra that were sensible given the medico-chemical context. The central focus of the study is yet the raw performance of the models without the use of any simulation data i.e. instance 2. In all instances, training was roughly 80\% of the data and 20\% left for testing.

\begin{figure}[h]
\centering
\includegraphics[width=\linewidth]{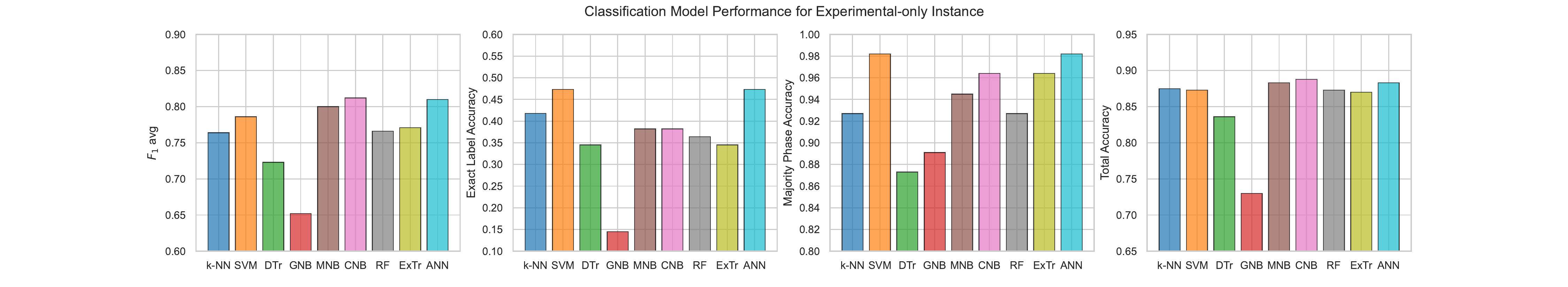}
\caption{Experimental-only Classification Performance}
\label{fig:bars}
\end{figure}

In Tables \ref{tab:identification} and \ref{tab:identification2}, the summarized results are displayed. Table \ref{tab:identification} contains model performances for the first two instances and Table \ref{tab:identification2} shows the last two. Figure \ref{fig:bars} depicts results for instance 2 alone.  We present the $F_1$ micro (\textit{mic}), weighted (wtd) and samples (smps) scores and their average (avg). $F_1$ macro scores were omitted as several phase labels are naturally less common than others [Table \ref{tab:ks}]. Accuracy is registered as three types: the exact-label accuracy (ELA),  the majority phase accuracy (MPA) and the total accuracy. The total accuracy is the standard definition which considers the overall correctness of all the predictions made by the classifier. The MPA is the percentage of samples whereby at least the majority phase composition is predicted positively. The ELA is the percentage of samples in the dataset where all phases are exactly predicted. That is to say if 3 of 7 possible phases exist in the sample, then these are predicted as `1' while all others predicted as `0'. In the case where one of the phases is not predicted or where an additional phase is predicted outside of the original 3, this is rendered as an incorrect prediction. As to be expected, ELA values were lower than the MPA and total accuracies as the qualification criterion is quite stringent. In the case where the entire experimental set was reserved for testing i.e. instance 3, the scores reflected are for a single evaluation. Otherwise, for instances 1 and 2, $F_1$ and accuracy scores were the average of 5-fold cross-validation and for instance 4 the average of 10 evaluations. 

The simulation-only evaluation i.e. instance 1, was split for the primary and secondary simulation datasets. The Gaussian Naive Bayes (GNB) classifier performed exceptionally with a total accuracy of 99\% ($F_1\; \text{avg}=\;$ 0.952, $\text{MPA}=\;$0.933) for monophasic spectra which implies a Gaussian spread of the characteristic features in each phase for simulated data. The Support Vector classifier (SVM) as well as the other Naive Bayes models also showed a strong performance alongside the shallow artificial neural network (ANN). The SVM attained the second highest $F_1\;$ average and total accuracy with 0.924 and 98.5\% respectively. The Complement Naive Bayes (CNB) method had the second best majority phase accuracy $\text{MPA}=\;$0.914. Although classifying single-phase spectra may not appear a major challenge, the simulation protocol produced significant alterations to peak intensity, position and widths for each class. The ability to still recognize a phase despite these large changes is noteworthy. Appending the multiphasic simulations, we see a drastic decrease for the GNB classifier but maintained performance of ANN, SVM and other Naive Bayes classifiers. In this case, the ANN scored highest with $F_1\; \text{avg}=\;$0.827 and a 93.2\% total accuracy, but SVM was a very close second with $F_1\; \text{avg}=\;$0.820 and 92.7\% total accuracy. The exact-label was also predicted quite well for both models with 56.8\% and 56.2\% for ANN and SVM accordingly. It may be said that SVM performed on par with the ANN despite requiring a substantially shorter training time, which is 0.08 seconds compared to 88 seconds for the ANN with this dataset. With regards to accurately identifying the majority phase, the other two Naive Bayes classifiers topped the SVM and ANN with $\text{MPA}=\;$0.968 for CNB and $\text{MPA}=\;$0.941 for Multinomial Naive Bayes (MNB). 

\begin{table}[t]
\caption{Model Performance for Simulation-only and Experimental-only Training and Testing.\\
\scriptsize{The left side shows data for instance 1 where models are trained and tested on simulations only and the right side for instance 2 with real, experimental data only. $F_1$ and accuracy scores are the average of 5-fold cross-validation. `ELA' stands for exact-label accuracy.  `MPA' is the majority phase accuracy. `DT' is the dataset type for simulations-only, i.e. only single-phase denoted `s' for which there were 105 spectra, or both single and multiphase denoted by `m' which consisted of 184 spectra. Values in parentheses (\;) show any improved performance if phases in quantities less than 10\% were ignored. Scores in \textbf{bold} highlight the two highest values for each metric and category.
\textbf{* k-NN} $k$-Nearest Neighbors, \textbf{SVM} Support Vector Machine, \textbf{DTr} Decision Tree, \textbf{GNB} Gaussian Naive Bayes, \textbf{MNB} Multinomial Naive Bayes, \textbf{CNB} Complement Naive Bayes, \textbf{RF} Randomized Forest, \textbf{ExTr} Extremely Randomized Trees, \textbf{ANN} Artificial Neural Network (Multi-Layer Perceptron)}
}
\label{tab:identification}
\vspace{2mm}
\begin{threeparttable}
\resizebox{\textwidth}{!}{%
\begin{tabular}{lrrrrrrrrrrrrrrrr}\toprule
\textbf{} &\multicolumn{8}{c}{\textbf{[1]\;\; Simulation-only Training \& Testing}} &\multicolumn{7}{c}{\textbf{[2]\;\;Experimental-only Training \& Testing}} \\\cmidrule{2-16}
\multirow{2}{*}{\textbf{Model*}} &\multirow{2}{*}{DT} &\multicolumn{4}{c}{\textbf{$F_1$ Scores}} &\multicolumn{3}{c}{\textbf{Accuracy}} &\multicolumn{4}{c}{\textbf{$F_1$ Scores}} &\multicolumn{3}{c}{\textbf{Accuracy}} \\\cmidrule{3-16}
& &mic &wtd &smps &\textbf{avg} &ELA &MPA &Total &mic &wtd &smps &\textbf{avg} &ELA &MPA &Total \\\midrule
\multirow{2}{*}{\textbf{k-NN}} &s &0.804 &0.765 &0.695 &0.755 &0.695 &0.695 &0.952 &\multirow{2}{*}{0.768 (0.792)} &\multirow{2}{*}{0.739 (0.766)} &\multirow{2}{*}{0.785 (0.798)} &\multirow{2}{*}{0.764} &\multirow{2}{*}{\textbf{0.418}} &\multirow{2}{*}{0.927} &\multirow{2}{*}{0.875 (0.894)} \\
&m &0.742 (0.749) &0.721 (0.731) &0.718 (0.729) &0.727 &0.416 (0.459) &0.827 &0.895 (0.900) & & & & & & & \\
\multirow{2}{*}{\textbf{SVM}} &s &\textbf{0.944} &\textbf{0.932} &0.895 &\textbf{0.924} &\textbf{0.895} &0.895 &\textbf{0.985} &\multirow{2}{*}{0.781 (0.796)} &\multirow{2}{*}{0.772 (0.800)} &\multirow{2}{*}{0.805 (0.813)} &\multirow{2}{*}{0.786} &\multirow{2}{*}{\textbf{0.473 (0.509)}} &\multirow{2}{*}{\textbf{0.982}} &\multirow{2}{*}{0.873 (0.886)} \\
&m &\textbf{0.828 (0.834)} &\textbf{0.824 (0.832)} &\textbf{0.809 (0.818)} &\textbf{0.820} &\textbf{0.562 (0.589)} &0.924 &\textbf{0.927 (0.931)} & & & & & & & \\
\multirow{2}{*}{\textbf{DTr}\tnote{a}} &s &0.629 &0.644 &0.629 &0.634 &0.629 &0.629 &0.894 &\multirow{2}{*}{0.681} &\multirow{2}{*}{0.680 (0.684)} &\multirow{2}{*}{0.691} &\multirow{2}{*}{0.684} &\multirow{2}{*}{0.291} &\multirow{2}{*}{0.818} &\multirow{2}{*}{0.810 (0.813)} \\
&m &0.579 &0.564 &0.562 &0.568 &0.276 (0.292) &0.616 &0.822 & & & & & & & \\
\multirow{2}{*}{\textbf{DTr}\tnote{b}} &s &0.638 &0.639 &0.638 &0.638 &0.638 &0.638 &0.897 &\multirow{2}{*}{0.716} &\multirow{2}{*}{0.718 (0.732)} &\multirow{2}{*}{0.736} &\multirow{2}{*}{0.723} &\multirow{2}{*}{0.345 (0.364)} &\multirow{2}{*}{0.873} &\multirow{2}{*}{0.836 (0.844)} \\
&m &0.631 &0.624 &0.616 (0.618) &0.624 &0.357 (0.384) &0.703 &0.842 (0.845) & & & & & & & \\
\multirow{2}{*}{\textbf{GNB}} &s &\textbf{0.964} &\textbf{0.959} &\textbf{0.933} &\textbf{0.952} &\textbf{0.933} &\textbf{0.933} &\textbf{0.990} &\multirow{2}{*}{0.625} &\multirow{2}{*}{0.714 (0.734)} &\multirow{2}{*}{0.617} &\multirow{2}{*}{0.652} &\multirow{2}{*}{0.145 (0.182)} &\multirow{2}{*}{0.891} &\multirow{2}{*}{0.730 (0.738)} \\
&m &0.658 &0.645 &0.653 &0.652 &0.297 &0.849 &0.821 & & & & & & & \\
\multirow{2}{*}{\textbf{MNB}} &s &0.923 &0.912 &0.863 &0.899 &0.857 &0.867 &0.98 &\multirow{2}{*}{0.795 (0.798)} &\multirow{2}{*}{\textbf{0.808 (0.828)}} &\multirow{2}{*}{0.796} &\multirow{2}{*}{0.8} &\multirow{2}{*}{0.382} &\multirow{2}{*}{0.945} &\multirow{2}{*}{\textbf{0.883 (0.891)}} \\
&m &0.818 &0.817 (0.819) &0.805 (0.806) &0.813 &0.503 (0.514) &\textbf{0.941} &0.917 & & & & & & & \\
\multirow{2}{*}{\textbf{CNB}} &s &0.910 &0.915 &\textbf{0.902} &0.909 &0.876 &\textbf{0.914} &0.974 &\multirow{2}{*}{\textbf{0.808 (0.811)}} &\multirow{2}{*}{\textbf{0.820 (0.842)}} &\multirow{2}{*}{\textbf{0.807}} &\multirow{2}{*}{\textbf{0.812}} &\multirow{2}{*}{0.382} &\multirow{2}{*}{\textbf{0.964}} &\multirow{2}{*}{\textbf{0.888 (0.896)}} \\
&m &0.810 &0.815 &0.803 &0.809 &0.443 &\textbf{0.968} &0.908 & & & & & & & \\
\multirow{2}{*}{\textbf{RF}\tnote{c}} &s &0.706 &0.632 &0.552 &0.63 &0.552 &0.552 &0.935 &\multirow{2}{*}{0.769 (0.782)} &\multirow{2}{*}{0.746 (0.770)} &\multirow{2}{*}{0.783 (0.786)} &\multirow{2}{*}{0.766} &\multirow{2}{*}{0.364} &\multirow{2}{*}{0.927} &\multirow{2}{*}{0.873 (0.886)} \\
&m &0.641 (0.642) &0.590 (0.591) &0.541 &0.591 &0.330 (0.362) &0.578 &0.879 (0.883) & & & & & & & \\
\multirow{2}{*}{\textbf{ExTr}\tnote{c}} &s &0.759 &0.723 &0.619 &0.700 &0.619 &0.619 &0.944 &\multirow{2}{*}{0.774 (0.788)} &\multirow{2}{*}{0.760 (0.787)} &\multirow{2}{*}{0.780 (0.784)} &\multirow{2}{*}{0.771} &\multirow{2}{*}{0.345 (0.364)} &\multirow{2}{*}{\textbf{0.964}} &\multirow{2}{*}{0.870 (0.883)} \\
&m &0.649 (0.654) &0.576 (0.580) &0.570 (0.572) &0.598 &0.330 (0.362) &0.589 &0.879 (0.884) & & & & & & & \\
\multirow{2}{*}{\textbf{ANN}} &s &0.938 &0.912 &0.886 &0.912 &0.886 &0.886 &0.984 &\multirow{2}{*}{\textbf{0.802 (0.808)}} &\multirow{2}{*}{0.807 (0.822)} &\multirow{2}{*}{\textbf{0.820}} &\multirow{2}{*}{\textbf{0.810}} &\multirow{2}{*}{\textbf{0.473 (0.509)}} &\multirow{2}{*}{\textbf{0.982}} &\multirow{2}{*}{\textbf{0.883 (0.891)}} \\
&m &\textbf{0.837 (0.839)} &\textbf{0.826 (0.831)} &\textbf{0.817 (0.824)} &\textbf{0.827} &\textbf{0.568 (0.595)} &0.924 &\textbf{0.932 (0.934)} & & & & & & & \\
\bottomrule
\end{tabular}
}
\begin{tablenotes} \scriptsize
\item[a] using gini optimization. 
\item[b] using entropy optimization.
\item[c] for [1] gini optimized with 70 estimators; for [2] entropy optimized with 70 estimators
\end{tablenotes}
\end{threeparttable}
\end{table}


\begin{table}[h]
\caption{Model Performance for Testing on Experimental Data using an Augmented Training Dataset.\\ \scriptsize{
The left side of the table reflects results from a single iteration of training with the secondary simulated dataset (N=184) comprising monophase and multi-phase synthetic spectra. The test samples were the complete experimental dataset (T=52). The right side of the table shows averaged results from 10 iterations of training with 105 simulated single-phase spectra and 21 randomly selected experimental scans. Testing was performed on the remainder T=31 experimental spectra. Values in parentheses (\;) show any improved performance if phases in quantities less than 10\% were ignored. Scores in \textbf{bold} highlight the two highest values for each metric and category.}}\label{tab:identification2}
\vspace{2mm}
\begin{threeparttable}
\resizebox{\textwidth}{!}{
\begin{tabular}{lrrrrrrrrrrrrrr}\toprule
\textbf{} &\multicolumn{7}{c}{\textbf{[3]\;\;Simulation-only Training for Experimental Testing}} &\multicolumn{7}{c}{\textbf{[4]\;\;Simulation + Experimental Training for Experimental Testing}} \\\cmidrule{2-15}
\multirow{2}{*}{\textbf{Model*}} &\multicolumn{4}{c}{\textbf{$F_1$ Scores}} &\multicolumn{3}{c}{\textbf{Accuracy}} &\multicolumn{4}{c}{\textbf{$F_1$ Scores}} &\multicolumn{3}{c}{\textbf{Accuracy}} \\\cmidrule{2-15}
&mic &wtd &smps &\textbf{avg} &ELA &MPA &Total &mic &wtd &smps &\textbf{avg} &ELA &MPA &Total \\\midrule

\textbf{k-NN} &0.72 (0.74) &0.68 (0.70) &0.76 (0.77) &0.72 &0.346 &0.923 &0.857 (0.874) &0.712 (0.734) &0.680 (0.700) &0.730 (0.742) &0.707 &0.348 &0.865 &0.855 (0.871) \\
\textbf{SVM} &\textbf{0.79} &\textbf{0.78 (0.79)} &\textbf{0.79} &\textbf{0.787} &0.346 &\textbf{1.000} &\textbf{0.879 (0.885)} &\textbf{0.806 (0.820)} &\textbf{0.791 (0.807)} &\textbf{0.822 (0.831)} &\textbf{0.806} &\textbf{0.484 (0.506)} &\textbf{0.984} &\textbf{0.888 (0.901)} \\
\textbf{DTr\tnote{a}} &0.60 &0.58 &0.61 &0.597 &0.192 &0.712 &0.755 &0.646 (0.648) &0.631 (0.637) &0.655 (0.659) &0.644 &0.297 (0.313) &0.758 &0.793 (0.804) \\
\textbf{GNB} &0.71 &0.71 (0.72) &0.71 &0.710 &0.365 (0.385) &0.865 &0.824 (0.835) &0.601 &0.632 &0.625 &0.619 &0.213 (0.229) &0.861 &0.726 \\
\textbf{MNB} &\textbf{0.78 (0.79)} &\textbf{0.77 (0.79)} &\textbf{0.78} &0.777 &\textbf{0.442 (0.481)} &0.942 &0.874 (0.885) &0.702 (0.741) &0.679 (0.722) &0.728 (0.756) &0.703 &0.397 (0.413) &0.916 &0.849 (0.876) \\
\textbf{CNB} &\textbf{0.79 (0.80)} &\textbf{0.78 (0.80)} &\textbf{0.78 (0.79)} &\textbf{0.783} &\textbf{0.423 (0.481)} &\textbf{0.962} &\textbf{0.876 (0.887)} &0.727 (0.758) &0.708 (0.744) &0.746 (0.769) &0.727 &0.406 (0.435) &0.929 &0.856 (0.879) \\
\textbf{RF\tnote{b}} &0.54 (0.55) &0.51 (0.52) &0.53 &0.527 &0.212 &0.635 &0.786 (0.802) &0.642 (0.671) &0.588 (0.618) &0.643 (0.660) &0.624 &0.326 (0.339) &0.774 &0.827 (0.851) \\
\textbf{ExTr\tnote{c}} &0.55 &0.50 &0.54 &0.530 &0.250 &0.577 &0.797 (0.808) &0.673 (0.690) &0.636 (0.657) &0.666 (0.679) &0.658 &0.329 &0.800 &0.829 (0.847) \\
\textbf{ANN} &0.76 (0.79) &0.75 (0.78) &\textbf{0.78 (0.80)} &0.763 &0.404 (0.423) &\textbf{1.000} &0.871 (0.893) &\textbf{0.809 (0.819)} &\textbf{0.794 (0.806)} &\textbf{0.825 (0.832)} &\textbf{0.809} &\textbf{0.494 (0.523)} &\textbf{0.984} &\textbf{0.889 (0.901)} \\
\bottomrule
\end{tabular}}
\begin{tablenotes} \scriptsize
  \item[a] entropy optimization
  \item[b] entropy optimization with 70 estimators
  \item[c] gini optimization with 70 estimators
\end{tablenotes}
\end{threeparttable}
\end{table}


For instance 2 which dealt solely with experimental spectra, all classifiers showed good ability to positively predict the majority phase. To be sure, the experimental dataset exhibited less varied features in each class label than the simulated data. Hence, models which performed relatively poorly on the synthetic spectra performed better on actual experimental data. For example, the tree-based classifiers gave the lowest majority phase accuracies in instance 1 between 57.8\% for Random Forest (RF) and 70.3\% for the entropy-optimized Decision Tree (DTr). These values rose to 87.3\% and 92.7\% in instance 2. The best MPA for instance 2 was 98.2\% attained by both SVM and ANN. The exact-label accuracy was also the highest for these two models with all phases being correctly assigned for 47.3\% of samples. In terms of overall performance, the CNB classifier attained the best $F_1\;$ average and total accuracy of $0.812$ and 88.8\% accordingly. This was followed very closely by the ANN with $F_1\; \text{avg}=\;$0.810 and 88.3\% total accuracy. It must again be mentioned that the training time for CNB was a much shorter 0.03 seconds to 67 seconds for the ANN. 

Table \ref{tab:identification2} records the data for instances 3 and 4 which tests exclusively on experimental data but integrates the synthetic profiles for training. For instance 3, the majority phase of all testing samples were perfectly predicted by the SVM and ANN. The next best was 96.2\% for the CNB classifier. SVM outperformed ANN with $F_1\; \text{avg}=\;$0.787 versus 0.763, and only by a little in total accuracy with 87.9\% versus 87.1\%. The Complement and Multinomial Naive Bayes' methods both also did better than the ANN in terms of $F_1\;$ average. The exact-label accuracy was highest for MNB and CNB with 44.2\% and 42.3\% respectively. Lastly, a fraction of the experimental data was subsumed in the primary simulation training set for instance 4. We observe the highest scoring by the ANN and SVM classifiers once more at $F_1 \; \text{avg}=\;$ 0.809 with 88.9\% total accuracy against $F_1 \;\text{avg}=\;$0.806 and 88.8\% accordingly. Second to these was CNB, though there was a solid margin between Naive Bayes and the former two. 

Results recorded for instances 2 to 4 are most relevant to practice. From these, traditional models gave the best results  collectively in instance 2 for MPA and $F_1$ score regardless of limited experimental training data. Expectantly, when synthetic data was supplemented for experimental data in training the collective performance  of the classifiers dropped. With both SVM and ANN, we note that the prediction accuracy of the majority phase solidified more with added synthetic training. This was not true for other models. We particularly observe that overall ANN and SVM gave the best result in instance 4. The case with the ANN was that though MPA increased from instance 2 to 3, $F_1 $ decreased implying that synthetic-only training made discerning secondary and minor phases more difficult.

In summary, tree-type classifiers consistently gave lower performances such that the Random Forest and Extremely Randomized Trees (ExTr) ensemble methods rarely provided any benefit over a single Decision Tree estimator. Only in instance 2 an advantageous margin developed for the ensembles in MPA, but still with median $F_1$ scores. The Gaussian Naive Bayes classifier typically showed the poorest performance when tested on real, multiphasic spectra. Nearest Neighbors (k-NN) had average performance with experimental-only data and ranked 4th in traditional classifiers under the MNB, CNB and SVM in instance 3 and 4.  The Support Vector Machine was the best overall traditional model which performed comparably with or even slightly superior to the shallow Artificial Neural Network, yet training in only hundredths of a second. Complement Naive Bayes was second to SVM. As a final note, the CNB and MNB models did perform better with experimental-only training but the SVM took over once the training set was increased. 

\subsection*{Estimation of Phase Weight Fractions - Multi-Output Regression}
\begin{figure}[h]
\centering
\includegraphics[width=\linewidth]{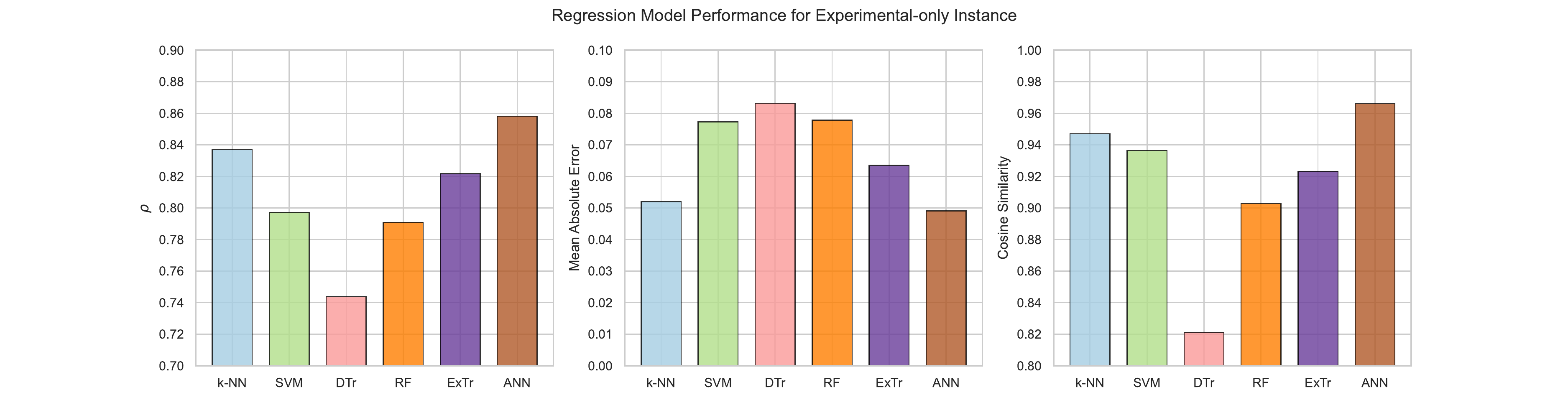}
\caption{Experimental-only Regressor Performance}
\label{fig:bars2}
\end{figure}

We replicated the same training-testing scenarios of the previous section for the prediction of phase fractions. Given a more restricted experimental dataset (N=46), we instead applied leave-one-out cross-validation (LOOCV) only for instance 2 where one sample at a time was reserved for testing while all others used in training. The results for mean absolute error \textbf{MAE}, cosine similarity \textbf{CS} and $\rho$ in each instance are recorded in Table \ref{tab:regress}. Instance 2 results alone are portrayed in Figure \ref{fig:bars2}. Upon inspection of the outputs, the ANN was one of two models that frequently returned negative weight predictions. The normalization requirement for the total phase fractions to sum to 1 was also not generally maintained. The SVM regressor had similar fault. Though not explicitly constrained, the tree-type and k-NN algorithms were able to reflect the positivity and normalized conditions for the output variables based on the training data seen. We enforced post-prediction constraints for ANN and SVR. Tabulated values reflect the performance metrics after these considerations. 

The shallow ANN and k-NN regressors stood out across all instances. The k-NN showed significantly better performance than all other models  in instances 1 and 3 with $\rho=\;$0.776, $\text{MAE}=\;$0.068 and $\rho=\;$0.846, $\text{MAE}=\;$0.048 accordingly. The ANN performed a little better ($\rho=\;$0.858, $\text{MAE}=\;$0.049) than SVM ($\rho=\;$0.837, $\text{MAE}=\;$0.052) in experimental-only instance 2. For instance 3, the scores were split as the k-NN model achieved the better mean absolute error (MAE$=\;$0.083) where the ANN had a better $\rho$ of 0.766. Granted, constraining outputs caused notable improvements for both the ANN and SVR models. Particularly with instance 3, the ANN mean absolute error decreased from 0.118 to 0.088 and cosine similarity increased from $\text{CS}=\;$0.876 to 0.885. With the Support Vector, $\text{MAE}=\;$0.180 decreased to 0.132 and $\text{CS}=\;$0.666 increased to 0.752.

Collectively, regression models performed best to worst in instance 2, 4, 3 then 1 with $\rho\; \text{avg}=\;$0.808, 0.805, 0.712 and 0.674 respectively. Despite only about three dozen training examples, instance 2 gave also the lowest averaged MAE of 0.067. Training only with simulated data i.e. instance 3, was not ideal even if all major experimental and sample effects were modelled. Inclusion of some experimental data in training saw some advantage over instance 2 for ensemble methods and k-NN alone.   

\begin{table}[h]\centering
\caption{Regression Model Performance for Phase Fraction Estimation  on Simulated and Experimental Datasets.\\ \scriptsize{
Instance [1] reports on 5-fold cross-validation on the simulated dataset (N=184) and instance [2] shows the average of leave-one-out cross validation on the experimental dataset (N=46). Instance [3] relates to testing on the experimental dataset while training only on the simulated dataset. Instance [4] uses all simulated data plus randomly selects half of the experimental data for training (N=207), then tests on the next half of the experimental (T=23). The averages of 10 evaluations corresponding to 10 randomly selected sets from the experimental data are registered. The highest score for each metric and instance is \textbf{bolded}.} $^\dag$ \textbf{k-NN} Nearest Neighbors regressor (k=2, distance weighted); \textbf{DTr} Decision Tree regressor (default); \textbf{RF} Random Forest regressor (default); \textbf{ExTr} Extremely \\ Randomized Trees (estimators=70); \textbf{ANN} Multi-Layer Perceptron regressor. * \textbf{SVR} Support Vector regressor with radial basis function kernel for instances [1] and [3-4] and with linear kernel  (c=1) for instance [2].}
\label{tab:regress}
\vspace{2mm}
 \begin{threeparttable}
\resizebox{\textwidth}{!}{%
\begin{tabular}{lcccccccccccc}\toprule
\multirow{2}{*}{\textbf{Model\tnote{$\dag$}}} &\multicolumn{3}{c}{\textbf{[1]\; Sim-only}} &\multicolumn{3}{c}{\textbf{[2]\; Exp-only}} &\multicolumn{3}{c}{\textbf{[3]\; Sim Training}} &\multicolumn{3}{c}{\textbf{[4]\; Sim+Exp Training}} \\\cmidrule{2-13}
&$\rho$ &\textbf{MAE} &\textbf{CS} &$\rho$ &\textbf{MAE} &\textbf{CS} &$\rho$ &\textbf{MAE} &\textbf{CS} &$\rho$ &\textbf{MAE} &\textbf{CS} \\\midrule
\textbf{k-NN } &\textbf{0.7756} &\textbf{0.0682} &0.8718 &0.8369 &0.0520 &0.9470 &0.7535 &\textbf{0.0827} &0.8565 &\textbf{0.8462} &\textbf{0.0479} &\textbf{0.9539} \\
\textbf{SVR\tnote{$*$}} &0.6830 &0.1195 &\textbf{0.8844} &0.7971 &0.0773 &0.9364 &0.7259 &0.1036 &0.8612 &0.7907 &0.0804 &0.9508 \\
\textbf{DTr} &0.5418 &0.1377 &0.5774 &0.7438 &0.0832 &0.8211 &0.6244 &0.1245 &0.6362 &0.7536 &0.0781 &0.8349 \\
\textbf{RF} &0.6393 &0.1300 &0.7863 &0.7908 &0.0778 &0.9029 &0.6855 &0.1221 &0.8122 &0.8024 &0.0737 &0.9157 \\
\textbf{ExTr} &0.6953 &0.1079 &0.8313 &0.8217 &0.0635 &0.9232 &0.7156 &0.1085 &0.8044 &0.8362 &0.0601 &0.9392 \\
\textbf{ANN} &0.7108 &0.1029 &0.8738 &\textbf{0.8582} &\textbf{0.0491} &\textbf{0.9662} &\textbf{0.7660} &0.0879 &\textbf{0.8845} &0.8036 &0.0702 &0.9358 \\
\bottomrule
\end{tabular}}
\end{threeparttable}
\end{table}


\section*{Discussion}
Supervised learning investigations for many computational materials analysis tasks including phase identification, predominantly involve the application of deep neural networks. Neural models try to directly \textit{learn} the input-output mapping for a dataset whereas traditional models optimize on the basis of some assumption, be it parametric, methodic or both. Parametric learning models assume some structure or property of the data before parameter optimization through training. For illustration, Linear Regression models presuppose a linear function between features and their output and then optimize on the gradient weights for this linear structure. With Naive Bayes classifiers, a distribution model is selected, e.g. gaussian, for evaluating the probabilities of each class given the input data features. For other traditional models the assumption is less about data structure and more on the approach to group or classify the data. Example, the Support Vector Machine aims to resolve some optimal hyperplane for separating the different class outputs. Any testing data is afterward compared to a set of reference vectors belonging to the different classes which define the boundaries of the plane. SVM may however be viewed as parametric with the use of linear or non-linear kernel functions to derive the hyperplane. The $k$-Nearest Neighbors algorithm takes a ``learn by example" approach where the method is simply to label the test point according to the assigned labels of the $k$ closest points used in training the model.

Traditional learning algorithms are much faster to train and interpret as the computational approach taken to deduce a representation of the input-output relationship is much more straight-forward. They are also extendable to very many applications, failing to perform only if the embedded assumptions do not adequately portray the real problem. In these events, opting for more complex learning models is quite reasonable but presents a different form of challenge. Choosing the neural network architecture is the first as it is well-known that there are no standardized rules for it. The experimenter must establish the structural hyper-parameters relating to the type of network, the depth of layers and number of nodes in each layer, before settling on learning hyper-parameters like activation functions, gradient optimization and learning rates. Finding the optimal network structure by training and evaluating outputs with multiple hyper-parameters is a task in itself.

The next difficulty is that training complex architectures with a considerable number of parameters requires `big' datasets to gain the performance advantage over traditional learners that is anticipated. Yet, it takes time to experimentally acquire materials data and analyze it for the assignment of labels. Large-scale experimental libraries of well-characterized materials data within a given application is typically unachievable. Consequently, materials research involving deep learning heavily depends on simulation-based training, for which real-world applicability is yet to be ascertained. 

In our study, XRD data augmentation served to gauge its applicability in addition to the generalizability of learning models by imposing more variant data features for each phase. Simulated powder spectra can be made to resemble closely to actual scans by modelling various specimen and instrumental effects. For example, finite resolutions of real-world measurement devices ensure that diffraction peaks have determinable widths rather than the infinitesimally narrow theoretical line spectra. Furthermore, peak positions may be varied non-uniformly by altering material parameters for lattice dimensions and macro-strain, peak intensities by including preferred orientation, and peak shapes by setting finite crystallite sizes for each phase. The height of the powdered material on the sample plate can also give rise to a uniform positive or negative angular shift in peak positions for the entire scan. The above described effects as well as the addition of random noise and realistic background functions were all included in data augmentation. It is worth pointing out that the level of effort expended in preparing simulated spectra is dependent on the aim of the investigation, but mostly is at the discretion of the experimenter. For most effective training, we imagine that utter consideration should be made in formulating the parameter bounds for synthetic spectra generation within the applied context, as opposed to overshooting the objective entirely. Still in the present work  the best collective classifier performances were achieved for instance 2 which reinforces that the use of experimental data for training is the ideal scenario, even with a small batch.

To sum up, traditional supervised learning algorithms are still tenable for automating tasks in analysis of materials data and ought to be considered first in similar lines of investigation. Whilst some models were not so effective for our biomedical application, others like the Support Vector, Naive Bayes and $k$-Nearest Neighbors contended with the shallow Artificial Neural Network. Other studies regarding XRD phase-fraction estimation have likewise reported the success of traditional models over neural networks, even deep learners trained with millions of synthetic spectra \cite{lee2021data,park2022application}. In the current study, we have still not explored dimensionality reduction, nor have we properly ventured into ensemble learning techniques like stacking and boosting for enhancing model performance. Suzuki et al. \cite{suzuki2020symmetry} reduced simulated XRD line spectra to just 11 features and achieved high accuracy in crystal system prediction with a tree-ensemble classifier. Bunn et al. \cite{bunn2015generalized} developed a supervised learning model using AdaBoost \cite{freund1997decision} for feature extraction from materials spectral data and subsequent phase identification. Further study along these lines may prove beneficial in conclusively demonstrating the full adequacy of traditional models for XRD phase identification and phase-fraction estimation tasks. 

\section*{Conclusion}
In the current investigation relating to biomedical materials analysis, we found that the Support Vector Machine (SVM) and Complement Naive Bayes (CNB) classifiers were successful in multi-phase identification with experimental and simulated XRD spectra. For four training-testing scenarios, SVM and CNB were comparable to (and in some instances better than) the Artificial Neural Network despite training in mere hundredths of a second where the ANN took well over a minute. Quantifying relative phase fractions was also successfully executed by a $k$-Nearest Neighbors regressor which performed distinctly better than the ANN regressor in some scenarios and in others only slightly inferior to it. We thus conclude that traditional machine learning models are yet viable choices for task automation in materials analysis applications. They also provide the further advantages of being more interpretable, much faster to train, not requiring enormous training datasets, and having a searchable hyper-parameter space for model tuning.

\bibliographystyle{plain}
\bibliography{sample}

\end{document}